\documentclass{optica-article}
\journal{opticajournal} 
\articletype{Research Article}
\usepackage{lineno}
\usepackage[utf8]{inputenc}
\usepackage{amsmath}
\usepackage{graphicx, float}
\usepackage{mathtools, bm}
\usepackage{hyperref}
\usepackage[shortlabels]{enumitem}
\usepackage[version=4]{mhchem}
\usepackage{tensor}
 \usepackage[titletoc,title]{appendix}
\usepackage{rotating}
\usepackage{tikz}
\usepackage{caption}
\usepackage{subcaption}
\usepackage{upgreek}
\usepackage{siunitx}

\begin{document}




\title{Vortex Propagation in Orbital Angular Momentum Beams and the Effects of a Limited Aperture}

\author{Ryan Husband,\authormark{1} Jessica Eastman,\authormark{1} Ryan J. Thomas,\authormark{1} Simon A. Haine,\authormark{1} Rhys H. Eagle,\authormark{1} John D. Close,\authormark{1} and Samuel Legge\authormark{1,*}}

\address{\authormark{1}Department of Quantum Science and Technology, 
Research School of Physics, Australian National University, Canberra, ACT 2601, Australia\\
\email{\authormark{*}samuel.legge@anu.edu.au}}
    \begin{abstract}
    \centering When generating light with orbital angular momentum by imprinting orbital phase onto a standard Gaussian beam, it is often assumed that the propagation of the generated spatial mode is a Laguerre-Gaussian. However, the true propagation of this beam in a realistic, aperture-limited optical system is non-trivial and has not been thoroughly explored in existing literature. We explore a numerical model that shows the development of an optical vortex mode, propagating from the plane of phase modulation, and the relation of these dynamics to the orbital phase factor $\ell$ and the spatial bandwidth of the optical system. The results of this model are compared to experimental data for beams with $\ell$ values 1, 2, 5, and 10 propagating through a range of spatial filters, with the described model showing agreement in the near field regime.
    \end{abstract}

\section*{Introduction}
Optical modes with helical phase fronts carry an intrinsic orbital angular momentum (OAM) on the optical axis, defined by the phase gradient \cite{intrinsic_extrinsic,classical_electrodynamics}. These modes are generally referred to as optical vortices and are commonly represented by the eigenmode solutions of the paraxial wave equation (PWE) in cylindrical coordinates, the standard Laguerre–Gaussian (sLG) modes \cite{seigman}. The unique properties of optical vortices are useful in a range of applications, including optical communications \cite{OAM_use_1}, on-chip single photon sources \cite{OAM_use_2}, spectroscopy \cite{OAM_use_3}, and applying torque onto particles, acting as an optical spanner
\cite{optical_spanner_3,optical_spanner_1,optical_spanner_2,LG_origin}. 
Common methods used to generate optical vortices from a standard Gaussian beam include orbital phase modulation with a spiral phase plate (SPP) \cite{phase_plate}, q-plate \cite{qplate}, forked diffraction grating \cite{Fresnel_and_Fraunhofer}, or spatial light modulator (SLM) \cite{SLM_use,OAM_like_me1,OAM_like_me2,OAM_like_me3}. After the orbital phase optic, the core of the optical vortex diverges as it propagates into the near field. This divergence is related to the incident Gaussian mode, $\ell$, and the spatial frequency filtering effects of the optical system in a non-trivial way.
\newline

\noindent Consider the case in which an experimentalist requires an optical mode consisting of an approximate Gaussian transverse amplitude profile with an orbital phase, localized at a specific point in space: these modes are so-called Gaussian vortex (GV) modes \cite{GV_named}. In practice, where an optical setup imposes a limited aperture on a beam due to the finite size of the optical elements, a GV mode will develop a vortex core comprised of an intensity null centered over the phase singularity, characteristic of a beam with OAM. Experimental issues can arise when the size of the vortex core becomes large or diverges quickly relative to the aperture size of the optical system. In this paper, we will experimentally characterise both the vortex core size and divergence, and then compare our results with numerical and analytical models.
\newline

\noindent The GV mode has potential applications where an approximately uniform optical intensity profile with an orbital phase is required. Besides optical communication use cases \cite{GV_use_case,GV_use_2}, GV modes have prospective roles in the field of quantum metrology, specifically in atom interferometers. In atom interferometry, large Gaussian beams are used to ensure a uniform light intensity across the region where atoms are located. This uniformity is crucial for achieving consistent atom-light interactions, which directly impacts the accuracy of the measurements \cite{tophat_interferometry}. Ring-trapped atom interferometers, such as those detailed in \cite{use_case_1,use_case_2,use_case_3,use_case_4}, use the optical orbital phase and atom-light interactions to measure rotation rates. Most optical vortices, like sLG modes, have large radial intensity gradients which can lead to non-uniform atom-light interactions and reduce the accuracy of the rotation measurement. A GV mode combines the desirable intensity uniformity of Gaussian beams with the orbital phase structure required for rotation measurements, offering a favorable alternative to conventional vortex beams for enhancing rotation sensitivity in ring-trap atom interferometer configurations. 
\newline

\noindent We investigate two generalized relationships between the core size of an optical vortex made through orbital phase modulation and the orbital phase winding number $\ell$, the low pass frequency filtering (LPF) of the optical setup, and the propagation distance of the mode. First, a generalized relationship between the vortex core radius, the LPF, and $\ell$ is established. Second, the propagation distance of the vortex core over which its size changes is measured and related to $\ell$. These relationships are applicable to all experiments generating optical vortices using phase modulation and are experimentally verified through comparison to a model.

\section*{Theory}

\begin{figure}[ht]
\begin{center}
  \includegraphics[scale=0.8]{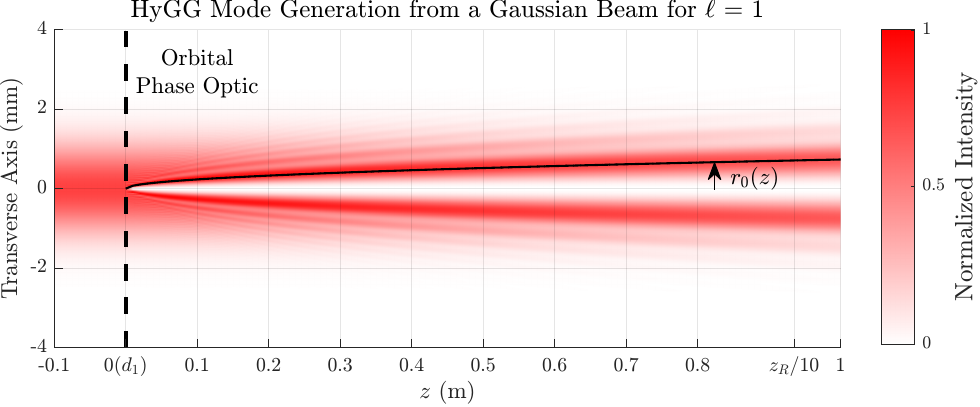}
  \caption{\centering Visual of a HyGG mode with $\ell = 1$ forming from orbital phase modulation of a \SI{780}{\nano\meter} wavelength Gaussian beam with a \SI{1.52}{\milli\meter} beam waist, produced by solving the PWE numerically with a GV mode as an initial condition at $d_1$. The primary ring radius, $r_0$, is shown as a function of the optical axis, $z$. The Rayleigh range, $z_\mathrm{R}$, defines the propagation distance where the vortex core diameter stabilizes to that of half of the beam diameter, $z_\mathrm{R}/10$ \cite{birth_and_evolution_of_optical_vortex}.}
  \label{fig:GV_propogation}
\end{center}
\end{figure}

\noindent The PWE is the standard model used for describing the propagation of an optical mode profile $\psi$ through free space along a central axis $z$, denoted by
\begin{equation}\label{eqn:PWE}
    2ik\frac{\partial\psi}{\partial z} + \nabla^2_{\perp}\psi = 0,
\end{equation}
\noindent for wavenumber $k$ and transverse Laplacian $\nabla^2_{\perp}$. The circular beam (CiB) is one of the basis solutions to the PWE, and provides a general representation of all forms of optical vortices and is thoroughly presented in the literature \cite{circle_beams,circle_beams2}. The CiB model has a high dimensionality that can be reduced to particular classes of optical vortices by specifying the separation constant $\gamma$, Seigman's complex beam parameter $q_0$, and integration constant $\Tilde{q}_0$ as derived in \cite{circle_beams}. For our work, we consider the family of Hypergeometric beams \cite{HyGG_fam} from the CiB basis as a practical modeling framework when a precise description of the vortex properties from phase-only modulation is required\cite{open_vortecies,vortex_book}. The hypergeometric beam family of includes the standard \cite{HyG} or auto-focusing \cite{autofocusing_HyGG} hypergeometric beams, Types I \cite{HyGG} and II \cite{HyGG_typeII} Hypergeometric Gaussian (HyGG), and the generalized HyGG \cite{HyGG_fam} beams. Type I HyGG modes, henceforth referred to as HyGG for brevity, form an over-complete non-orthogonal set of eigenmode solutions to the PWE and are an effective basis to model the orbital phase modulation of a fundamental Gaussian beam and its propagation into the near field \cite{Fresnel_and_Fraunhofer,F&F2,Kummer_beams,role_of_bw} as depicted in Fig. \ref{fig:GV_propogation}. 
\newline

\noindent Deriving the HyGG class from the CiB model requires the physical beam properties, including the position of the beam waist of the Gaussian component, $d_0$, the point of phase imprinting on the optical axis, $d_1$, and the Rayleigh range $z_\mathrm{R}$. The HyGG basis is derived in Appendix \ref{app:CiB_to_HyGG}, and is shown to approximate a GV mode in the limit of the phase imprinting plane in Appendix \ref{app:GV_derivation}

\begin{equation}\label{eqn:GV}
\begin{split}
\lim_{z \to d_1}\text{HyGG}^{\ell}_{p=-|\ell|}(r,\theta,z) &\propto \text{G}(r,q)e^{i\ell\theta},\\
\end{split}
\end{equation}

\noindent where $r$, $\theta$, and $z$ are the cylindrical coordinates, $p\in\mathbb{C}$ is an analogue of the radial quantum number for sLG modes \cite{circle_beams2}, and $\text{G}(r,q)$ is a standard Gaussian mode with complex beam parameter $q$ \cite{seigman}. Equation \ref{eqn:GV} demonstrates that the GV mode formed from imprinting a spiral phase in the thin-element approximation \cite{fourier_optics} only exists in a narrow region of $z$ near the phase imprinting plane; diffraction of the mode then leads to the formation of a vortex core for $z > d_1$ \cite{role_of_bw}. In principle, this GV mode can then be re-imaged at a different plane where the OAM and smooth radial intensity profile are useful.
\newline

\noindent In practice, all experimental generation of GV modes have limited aperture effects due to the finite size of the optical elements used in their creation. This paper characterises GV modes under these realistic conditions and explores the effect of this finite aperture by using a spatial LPF in the Fourier plane. When filtering a HyGG mode within the limit of $d_1$, the transverse intensity profile is morphed from a Gaussian distribution to a Gaussian-like spread with a noticeable intensity null correlating with the vortex core. Furthermore, the size of the vortex core remains roughly constant for a finite distance of the optical axis around $d_1$. In the same way that $z_\mathrm{R}$ is defined as the location where the beam waist $w_0(z)$ increases by $\sqrt{2}w_0(0)$ for a standard Gaussian beam, we define the uniformity distance of the vortex core, $z_\mathrm{V}$, as the distance the primary ring radius $r_0(z)$ increases by $\sqrt{2}r_0(0)$. As a practical definition, we define the GV mode as the aperture limited HyGG beam that is within a range $z_\mathrm{V}$ of $d_1$.
\newline

\noindent A simple approach to characterising a GV mode under a limited aperture is to identify a property of interest and numerically solve the PWE. We use this method to find $r_0(z)$ under a range of LPF conditions and $\ell$. We can experimentally measure $r_0(z)$ using a 4f imaging system (this is a Fourier filter, detailed in \cite{fourier_optics}) with a LPF on the central focal point, to find a relationship between $\ell$ and the LPF that is applicable to a range of similar optical systems.
\newline

\begin{figure}[H]
\begin{center}
\includegraphics[scale=1]{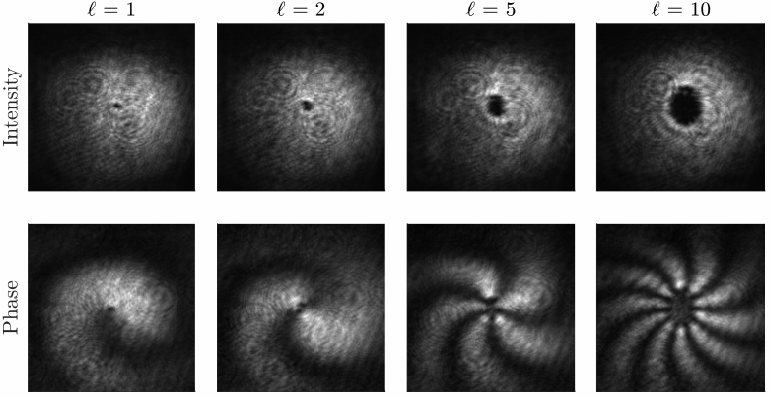}
  \caption{\centering The top row depicts intensity images captured of GV modes across several indicated $\ell$ values at the $d_1$ position on the optical axis. The bottom row depicts the corresponding phase profiles, generated through interference with a diverging Gaussian beam which adds a radial phase curvature and gives rise to the spiral phase patterns. These images were taken using the setup shown in Fig. \ref{fig:setup}.}
  \label{fig:GV_examples}
\end{center}
\end{figure}

\noindent Examples of experimentally measured GV intensity and phase profiles across several $\ell$ are shown from our experiment in Fig. \ref{fig:GV_examples} in the absence of the LPF. A noteworthy feature of the mode profiles in each of these cases is the vortex core present at all positions along the optical axis, including the phase imprinting plane. The core size also increases rapidly with $\ell$, visually demonstrating how a lack of understanding of the vortex core properties of a GV mode can be problematic in practical uses. Experimental realization of GV modes has been reported in the literature before, notably by Vallone in \cite{birth_and_evolution_of_optical_vortex}. Vallone highlights that the intensity null of a GV mode is an artifact of the limited aperture of an optical setup. Our work experimentally characterises the vortex core size at the phase imprinting plane as a function of a LPF size and $\ell$, and demonstrates how the size changes along the optical axis.

\noindent Our model of this optical system used the GV expression from Eq. \ref{eqn:GV} as the initial condition for the PWE at the phase imprinting plane. If $\ell$ is sufficiently large, the orbital spatial frequency of the optical mode near the phase singularity exceeds the assumptions underlying the PWE. Studies on large OAM sLG beams \cite{beyond_PWE,beyond_PWE2} suggest that for GV modes on the \SI{1}{\milli\meter} length scale we use, the upper bound on $\ell$ is $\mathord{\sim} 10^6$, confirming our values of $\ell$ are within the PWE regime. The PWE was then solved numerically at discrete positions along the optical axis to calculate the amplitude and phase profiles of the optical mode.  Numerically modeling the propagation of a GV mode enables a smooth transition from the ansatz of Eq. \ref{eqn:GV} into a paraxially valid mode, and the ability to match the measured mode profiles at $d_1$ to include experimental artifacts, such as the vortex core displacement from the center of the Gaussian profile, and the minor amplitude modulation from the phase imprinting method. Alternatively, analytical modeling of the system leads to expressions that cannot be resolved in closed form. For example, as we analytically derive in Appendix \ref{app:low pass filter derivation}, a GV mode under the influence of a LPF is denoted as
\begin{equation}\label{eqn:analytic_filtered_GV}
\begin{split}
\text{GV}_{\rm filt}(r,\theta)
&=Ce^{\,i\ell\theta}\int_{0}^{k_c}k_{r}^{\ell+1}U\!\Bigl(\tfrac{\ell+2}{2},\,\ell+1,\tfrac{\beta\,k_{r}^{2}}{4}\Bigr)
\,J_{\ell}(k_r r)\mathrm{d}k_r, \\
\end{split}
\end{equation}
for constants $C$, transverse radial wavenumber $k_r$, Tricomi’s confluent hypergeometric function $U$ \cite{Tricomi1955}, $\beta = w_{0}^{2}-2id_1/k$ for beam waist $w_0$, and Bessel function of the first kind $J_{\ell}$ of order $\ell$. While Eq. \ref{eqn:analytic_filtered_GV} cannot be expressed in a concise algebraic form, useful information regarding $r_0$ can still be obtained by solving
\begin{equation}\label{eqn:derivative}
\left.\frac{\partial}{\partial r}\bigl|\text{GV}_{\rm filt}(r,\theta)\bigr|\right|_{r=r_{0}} = 0,
\end{equation}
\noindent which leads to a transcendental equation. By introducing the dimensionless parameters $u = k_{c}r$ and $s = k_r r$, where $k_c=4\pi /D\lambda f$ for wavelength $\lambda$, LPF diameter $D$, and focal length of the objective lens in the Fourier filter $f$, a dimensionless radial profile function can be defined as
\begin{equation}\label{eqn:A_l}
    A(u) = \int_{0}^{u}s^{\ell+1}U\left(\frac{\ell+2}{2},\ell+1,\frac{\beta k_c^2 s^2}{u^2}\right)J_{\ell}(s)\mathrm{d}s,
\end{equation}
\noindent meaning Eq. \ref{eqn:analytic_filtered_GV} becomes $\text{GV}_{\rm filt}(r,\theta) = Cr^{-(\ell+2)}e^{i\ell\theta}A(u)$. We define $\alpha = r_0k_c$ as the first root of Eq. \ref{eqn:derivative}. There is no closed-form expression for the exact relationship between $\alpha$ and $\ell$; however, we can deduce an approximate relationship by solving Eq. \ref{eqn:A_l} numerically for a range of $\ell$, or by using analytical approximations as shown in Appendix \ref{app: solving A_l}. Regardless, one finds an approximate equivalence between $\alpha$ and the first positive root of $J_\ell$, denoted as $j_{\ell,1}$, producing the linear relationship $j_{\ell,1} \approx \alpha = a_1\ell + a_2$ where $a_1,a_2\in\mathbb{R}$ depend on the optical circuit. The physical core radius is therefore related to $\ell$ by
\begin{equation}\label{eqn:r_0_and_OAM}
r_0 = \frac{2f}{kD}(a_1\ell + a_2).
\end{equation}
\noindent The linear relationship between $r_0$ and the experiment's physical properties enables a simple way of relating $r_0$, $\ell$, $k$, and $D$ for a system using GV modes, assuming $a_1$ and $a_2$ are known. Therefore $\alpha$ can be considered a calibration factor for an optical circuit. 


\section*{The Experiment}
\begin{figure}[H]
\begin{center}
  \includegraphics[scale=0.54]{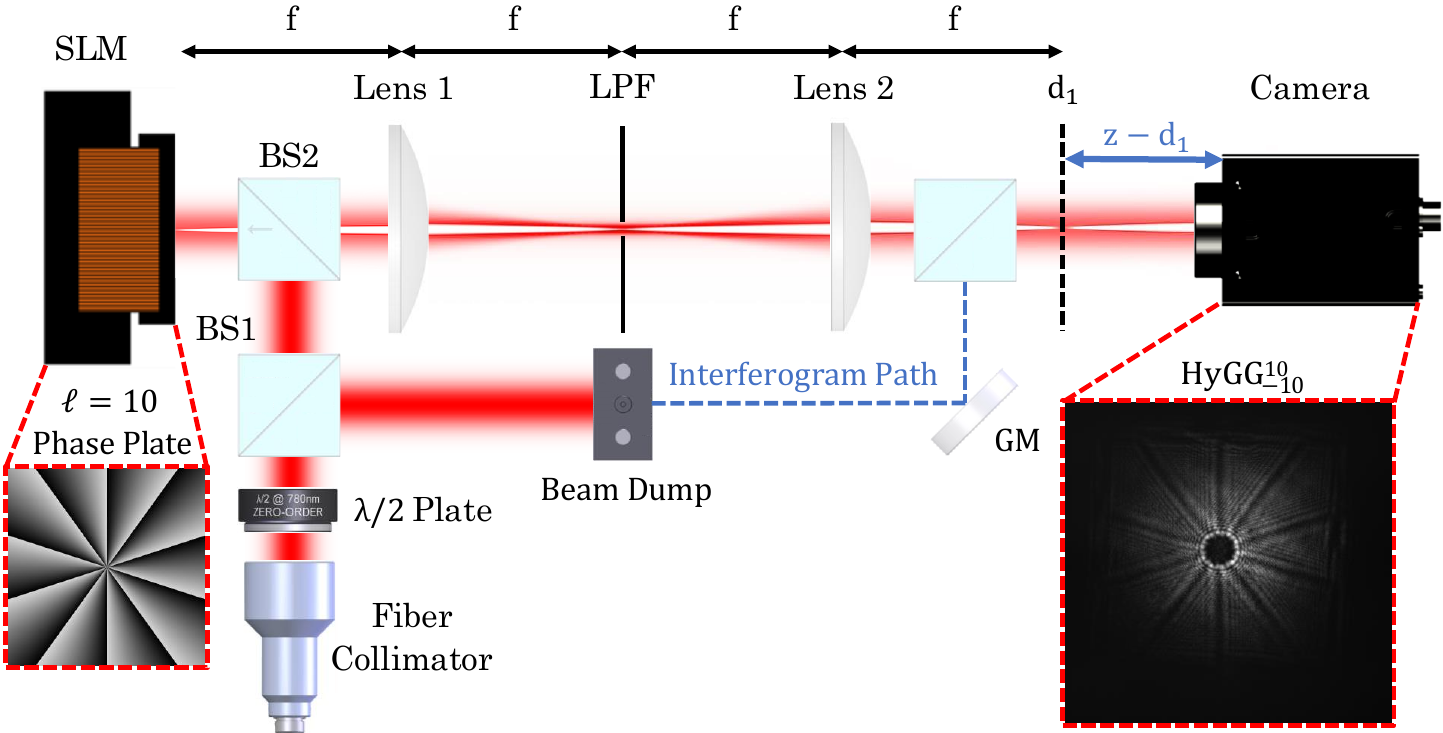}
  \caption{\centering Diagram of the setup as described in the text, where a $\text{HyGG}^{10}_{-10}$ mode is generated as an example SPP on the SLM, and intensity image on the camera.}
  \label{fig:setup}
\end{center}
\end{figure}

To validate the model described above, we built an experiment capable of generating GV beams from an SLM as shown in Fig. \ref{fig:setup}. This setup uses the aforementioned 4f imaging system to enable imaging of the beam in the generation plane of the orbital phase optic and at arbitrary distances after. It also allows verification of the orbital phase through an optional interference path. A \SI{780}{\nano\meter} Gaussian beam, with $w_0 =$ \SI{1.52}{\milli\meter}, emerges into free space from a single mode fiber coupled to a collimating output coupler. The beam is then polarization aligned for a Meadowlark SLM, model ODPDM512-0785 \cite{SLM_used}, with a half wave ($\uplambda/2$) plate, before passing through the first non-polarising beam splitter (BS1). This beam splitter separates the Gaussian beam into a transmission path to become a HyGG mode, and a reflective path to remain a Gaussian and traverse the indicated interferogram path if the beam dump is removed, where a gold mirror (GM) is used for interferogram alignment. 

\noindent The transmitted Gaussian from BS1 is then incident on the second non-polarising beam splitter, BS2, where the reflected mode goes to the SLM. The SLM provides a programmable means of phase modulating an incident beam through a liquid crystal display \cite{SLM_basics}. The SLM was calibrated to match the bit-depth to the appropriate optical phase modulation for optical vortex generation. The SLM was set as a SPP, such that the reflected beam converted into a HyGG beam with a set $\ell$ from the incident Gaussian beam. After emerging from the transmission port of BS2, the HyGG mode traverses a 4f imaging system, where each lens is a plano-convex, \SI{100}{\milli\meter} focal length lens. Each lens was \SI{50.8}{\milli\meter} in diameter to ensure the limiting aperture they added to the system was minimal compared to the effect of the LPF. The 4f imaging system creates a projection of the GV mode created by the SLM at the imaging plane $d_1$, located \SI{1.81}{\meter} from $d_0$. A CMOS camera was used to capture the GV mode and its divergence. The camera was translated along the optical axis ($z$) where intensity images of the optical mode were captured at 5 different locations $z$, from $d_1$, up to \SI{93}{\centi\meter} away. This distance range accounts for the birth of the vortex core until its stabilisation distance of $z_\mathrm{R}/10$ from Fig.~\ref{fig:GV_propogation}. To change the system's spatial frequency in a measurable way, 12 spatial LPFs were laser-cut from opaque acrylic disks with a \SI{25}{\milli\meter} outer diameter and a thickness of \SI{2}{\milli\meter}. Each disk had a central hole laser cut with diameters ranging from \SI{0.2}{\milli\meter} to \SI{5}{\milli\meter} for the beam to transmit through. When using a LPF, the disk was positioned on the focal point between lens 1 and lens 2 at the LPF plane indicated in Fig.~\ref{fig:setup} and was laterally centered on the optical vortex using a micrometer translation stage \cite{translation_stage}. The alignment was conducted using the camera located at $d_1$ in video mode, where the hole of the GV mode was made symmetric, indicating the LPF was centered on the vortex core.
\newline

\noindent The beam's orbital phase factor $\ell$ was visually confirmed on the camera after unblocking the interferogram path, and matched the $\ell$ used for the SPP. We quantified the mode composition using the procedure from \cite{reference}, allowing us to calculate the contribution of each mode in the sLG basis. The results revealed purities between $83-91\%$ in the $\ell$ set on the SPP, matching the results from comparable experiments which range from $77-93\%$ \cite{comparing_purities}.
\newline

\begin{figure}[ht]
\begin{center}
  \includegraphics[scale=0.8]{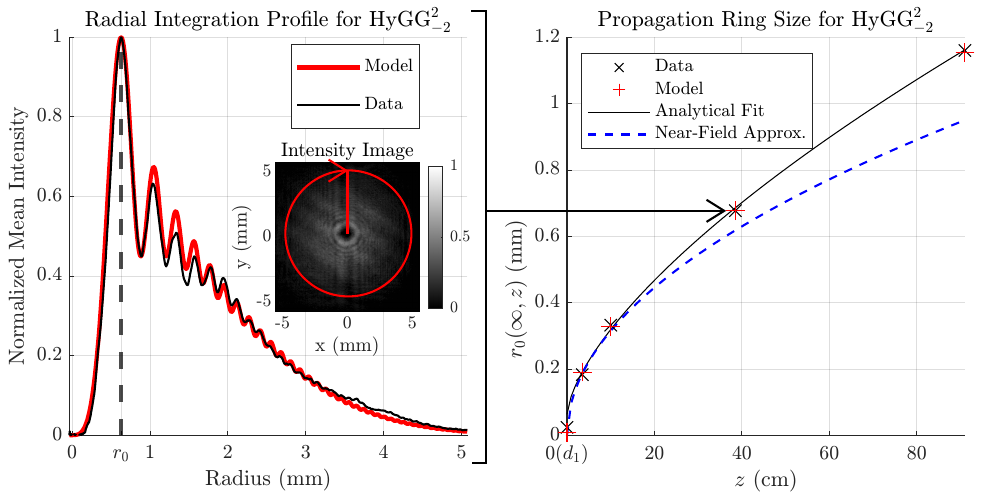}
  \caption{\centering \centering (Left) A comparison between the data (integrated orbitally from an intensity image as shown in the insert) and numerical model for $\text{HyGG}^2_{-2}$ at \SI{38.5}{\centi\meter} from $d_1$ with no LPF ($D = \infty$) applied, where $r_0(D,z)$ was measured. (Right) The $r_0(D,z)$ measurement was repeated for 5 different points along the optical axis and compared to the numerical model, the parametric fit from Eq. \ref{eq:Y2}, and the near field ring radius approximation from Eq.~28 of \cite{Fresnel_and_Fraunhofer}.}
  \label{fig:processed_data}
\end{center}
\end{figure}

\noindent We collected a series of intensity profiles of the GV modes along $z$ at the aforementioned camera locations, for each LPF disk, and for $\ell$ values of 1, 2, 5, and 10. A measurement with no LPF was also conducted for comparison to the literature. For each recorded intensity image, a dark image was captured for subtraction. An example of the intensity profile for $\ell = 2$ is shown in the left panel of Fig. \ref{fig:processed_data}. The center of the vortex core was manually identified on the image, and the image was integrated in the orbital direction to obtain the radial profile of the mode. The orbital integration process significantly increased the SNR of the radial profile by averaging out the artifacts on the intensity images from the experiment setup, such as the amplitude modulation from the SLM at the phase wrapping points of the SPP. For appropriate comparison of the mode profiles generated experimentally and in the numeric model, the orbital integration process was used in both instances. 

\section*{Analysis}

\noindent Using the radial profile of each optical mode, $r_0$ was measured as a function of $z$ and LPF diameter $D$ for each $\ell$ set on the SPP. An example of the measured $r_0$ along $z$ for the no-LPF case and $\ell = 2$ is shown in the right panel of Fig. \ref{fig:processed_data}. For $z>0$, the primary ring radius follows the expansion of a HyGG beam in the near field regime as approximated by Eq.~28 of \cite{Fresnel_and_Fraunhofer}, but deviates for increasing $z$. Notably, this deviation occurred for all $D$ and $\ell$ used in our experiment. For each $\ell$, a surface function $r_{0,\ell}(D,z)$ was fit to the ring radii data across our range of LPF diameters, $D$, and optical axis, $z$. To derive $r_{0,\ell}(D,z)$, we considered the mode divergence for the maximum and minimum LPF cases. The smallest LPF diameter used was $D =$ \SI{0.2}{\milli\meter}, and described the ring radii expansion across the optical axis for each $\ell$ with the function
\begin{equation}\label{eqn:use_in_fig_4}
r_{0,\ell}\left(0.2,z\right)=C_1+C_2\left(\sqrt{z^2+C_3^2}-C_3\right),
\end{equation}
where $C_1$ is the baseline for $r_0$ at $d_1$, $C_2$ sets the slope for large $z$ and $C_3>0$ controls the curvature near $z = d_1$. Alternatively, for the no-LPF case ($D = \infty$), the following power law best characterised the data for each $\ell$
\begin{equation}
r_{0,\ell}(\infty,z) = C_4 + C_5z^{C_6},
\label{eq:Y2}
\end{equation}
with vertical offset $C_4$, scale $C_5$, and exponent $C_6>0$. Notably $C_6\approx0.5$ for all $\ell$, similar to the ring radius approximation from \cite{Fresnel_and_Fraunhofer}. To capture the transition of $r_{0,\ell}(D,z)$ from $D =$ \SI{0.2}{\milli\meter} to $D = \infty$, we use the data at $z=0$ where $r_0$ is most sensitive to changes in $D$. Under these conditions, we find the weighting function as an appropriate fit
\begin{equation}
W(D)=\left(\frac{D}{D_0}\right)^{C_7},
\label{eq:weighting_function}
\end{equation}
where $D_0$ is the normalisation length of $W(D)$, and exponent $C_7<0$. Combining these functions results in the expression
\begin{equation}\label{eqn:surface_function}
r_{0,\ell}(z,D)=\left[1-W(D)\right]r_{0,\ell}(0.2,z) + W(D)r_{0,\ell}(\infty,z).
\end{equation}

\noindent Using Eq. \ref{eqn:surface_function}, we measure $z_\mathrm{V}$ as a function of $D$ for each $\ell$ data set to find a simple relationship for diverging GV modes as shown in Fig. \ref{fig:hole_lifetime}.


\section*{Results and Discussion}



\begin{figure}[H]
\begin{center}
  \includegraphics[scale=0.89]{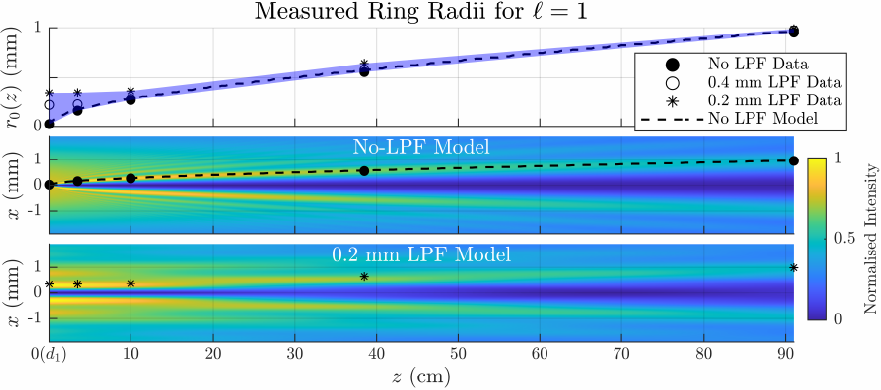}
  \includegraphics[scale=0.89]{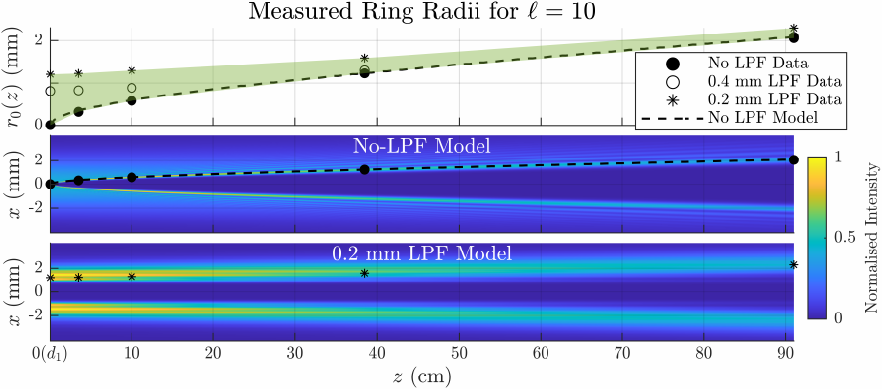}
  \caption{\centering Ring radii data sets and corresponding numerical model outputs for $\ell = 1$ and $10$ under $D = \SI{0.2}{\milli\meter}$ and when no LPF is applied, alongside a direct comparison to the numerical model for each. The shaded region for each $\ell$ contains the other 11 data sets for each LPF used, which are omitted from these plots except for the $D = \SI{0.4}{\milli\meter}$ case. The respective panels for the $\ell = 2$ and $\ell = 5$ datasets were excluded for readability.}
  \label{fig:Maximum Ring Radii Ranges}
\end{center}
\end{figure}

\noindent Our experiment offers a characterisation of the key properties that should be considered when using GV modes in an experiment, including the relationship between the radius of the mode's primary ring at the phase imprinting plane and the limited aperture of the preceding optical circuit, and how the ring radius diverges as the mode travels. The data recorded for $\ell = 1$ and $\ell = 10$ from this experiment are presented in Fig.~\ref{fig:Maximum Ring Radii Ranges}, alongside a comparison to the numerical model. 

\noindent At $d_1$ the intensity and interferogram images reveal an approximate Gaussian intensity profile when no LPF is applied, with a helical phase structure associated with the SPP index $\ell$, as predicted by Eq. \ref{eqn:GV}. When $D\leq \SI{4}{\milli\meter}$, a clear vortex core is revealed at $d_1$. The core size increases with larger $\ell$ and smaller $D$ near $d_1$. Further along the propagation axis, the ring radius always appears to approach the radii measured in the no-LPF case, likely due to the lower spatial frequency components of the mode at $d_1$ dominating the expansion rate for large $z$. 
\newline 

\noindent The numerical model of the experiment aligns with the data in the absence of a LPF, but its accuracy deteriorates with decreasing LPF size. As seen for $\ell = 10$ in Fig. \ref{fig:Maximum Ring Radii Ranges}, when comparing the experimentally measured values of $r_{0,\ell}(0.2,z)$ to those numerically calculated, the data matches the general behavior of the model, supporting the use of this model in verifying the data. In contrast, the data does not reflect the higher frequency structure that the numerical model predicts along $z$ due to the limited number of data points. The model implies that the LPF induces multiple radial peaks on the same primary ring as the mode propagates, which follow a staircase function when $r_{0,\ell}(0.2,z)$ is measured. This staircase function is not an artifact of the discretizations or field size of the optical modes used in the numerical model. Rather, this highlights that directly measuring the maximum transverse mode intensity as the primary ring radius is not the optimal approach to characterising the GV divergence under sufficiently small $D$ in the near field.
\newline

\begin{figure}[H]
        \includegraphics[width=\linewidth]{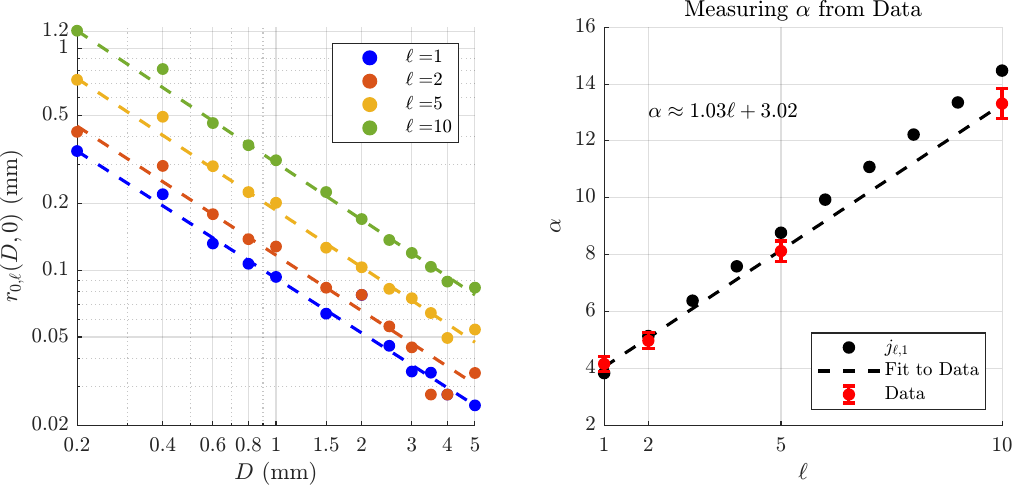}
        \label{fig:sub1}
    \caption{\centering (Left) The vortex core size at the phase imprinting plane, as measured across each $\ell$ and LPF. (Right) The mean $\alpha$ for each $\ell$, overlaid by the line of best fit.}
    \label{fig:alpha_l}
\end{figure}

\noindent The ring radii measured at $d_1$ for all LPF used, $r_{0,\ell}(D,0)$, are shown in the left panel of Fig. \ref{fig:alpha_l}. For each $\ell$ set on the SPP, the $n$-th measured ring radii from Fig.~\ref{fig:alpha_l} were related to $\alpha_{n,\ell}$ where
\begin{equation}
     \alpha_{n,\ell} =\frac{2\pi}{\lambda f}r_{0,\ell}(D_{n},0)D_{n}, 
\end{equation}
\noindent for $D_{n}$ as the $n$-th LPF diameter. All $\{\ell,\alpha\}$ data points were then linearly regressed to obtain $\alpha(\ell)$ for our experiment. A linear relationship between $\alpha$ and $\ell$ was measured, as depicted in the right panel of Fig.~\ref{fig:alpha_l}. Given that $\alpha$ is asymptotically equivalent to the first root of the Bessel function of order $\ell$, these values are included in Fig.~\ref{fig:alpha_l} for $\ell = [1,...,10]$ and follow the trend line $j_{\ell,1} = 1.177\ell + 2.805$. Similarly, the measured values for $\alpha$ follow the function $\alpha(\ell) = 1.03\ell + 3.02$. Although small, the differences between $j_{\ell,1}$ and $\alpha(\ell)$ are likely due to the assumptions from the derivation in Appendix \ref{app: solving A_l} showing that $r_{0,\ell}(D,0) = \alpha(\ell) \approx j_{\ell,1}$. Furthermore, unlike $j_{\ell,1}$, experiment dependent factors change $\alpha$, such as the uncertainty along $z$ when positioning the LPF on the Fourier plain. We therefore call $\alpha$ a calibration factor to relate $r_0$ to $\ell$ using Eq.~\ref{eqn:r_0_and_OAM}.

\begin{figure}[H]
    \centering
        \begin{subfigure}[b]{0.45\textwidth}
        \centering
        \includegraphics[width=\linewidth]{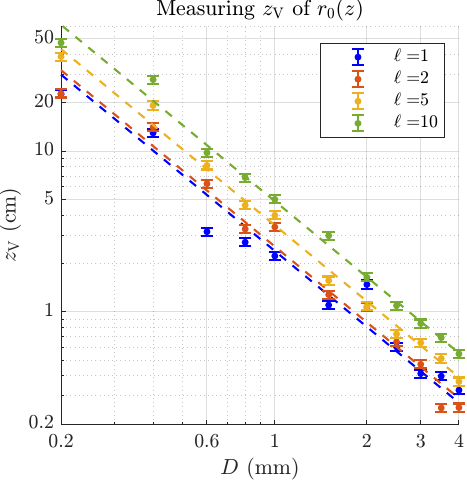}
    \end{subfigure}
    \hfill 
    \begin{subfigure}[b]{0.45\textwidth}
        \centering
        \includegraphics[width=\linewidth]{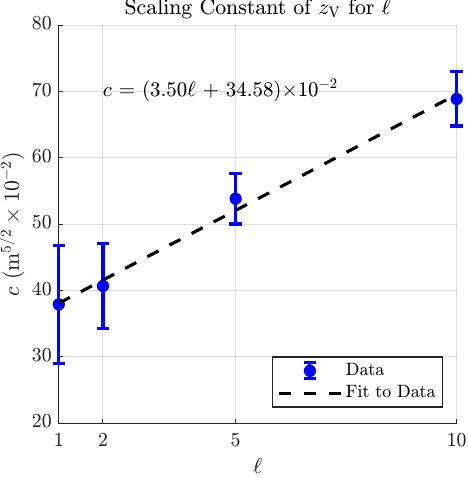}
    \end{subfigure}
    \caption{\centering (left) Demonstrating $z_\mathrm{V}$ as a function of $D$ per $\ell$ where the uncertainty of each data point is defined by the 95\% confidence bounds by fitting Eq. \ref{eqn:surface_function}. Fitting Eq. \ref{eqn:power_law} generated (right) the linear relationship between $\ell$ and $c$, scaled by $10^2$, with an uncertainty defined by the 95\% confidence bounds of the fit.}
    \label{fig:hole_lifetime}
\end{figure}

\noindent Moving from $d_1$ into the near field regime, $z_\mathrm{V}$ was measured for each $\ell$ by matching Eq.~\ref{eqn:surface_function} to the primary ring radii dataset by jointly fitting $C_1,...,C_7$ for every $(z,D)$. For each $D$, $z_\mathrm{V}$ was calculated as $r_{0,\ell}(D,z_\mathrm{V}) = \sqrt{2}r_{0,\ell}(D,0)$, demonstrating the following power law coupling between $D$ and $z_\mathrm{V}$ independent of $\ell$
\begin{equation}\label{eqn:power_law}
    z_\mathrm{V} = \frac{c}{D^\frac{3}{2}}
\end{equation}
for arbitrary $c\in\mathbb{R}$. Using a power law ensures that in the limit of $z_\mathrm{V}$ approaching zero, $D$ will tend towards infinity, as is observed in practice \cite{birth_and_evolution_of_optical_vortex,GV_named}. We measured $c$ for our optical system by performing a linear regression of the data in the left panel of Fig. \ref{fig:hole_lifetime}. The data corresponding to the $D = \SI{5}{\milli\meter}$ was excluded because $r_{0,\ell}(5,z_\mathrm{V})$ was below a measurable threshold. Notably, the $3/2$ power of $D$ was a best fit parameter for each dataset independent of $\ell$. Alternatively, the multiplicative factor $c$ appeared to change linearly with $\ell$ as demonstrated in the right panel of Fig. \ref{fig:hole_lifetime}. 

\noindent For a standard Gaussian beam, a larger beam waist allows the mode to retain a size comparable to that at the waist over a longer propagation distance, characterised by $z_\mathrm{R}$. Likewise, a larger vortex core at the phase imprinting plain,  $r_{0,\ell}(D,0)$, which increases with smaller $D$ and larger $\ell$, increases $z_\mathrm{V}$. It is uncertain whether the linear trend between $z_\mathrm{V}$ and $\ell$ is broadly applicable to standard optical circuits like $\alpha$ and $\ell$, or if $c$ can be considered an experimentally dependent calibration constant.
\newline

\noindent The behavior of the vortex core size at $d_1$ and along $z$ under different aperture limits demonstrates the considerations required to use GV modes in a practical setting. When the aperture is large, the mode at $d_1$ is indistinguishable from a Gaussian mode with an orbital phase. In this case, a GV mode would exist for a very small $z_\mathrm{V}$. Depending on the application, the user may need to ensure $z_\mathrm{V}$ is sufficiently large. From Fig.~\ref{fig:hole_lifetime}, applying a LPF will increase $z_\mathrm{V}$ at the cost of increasing $r_0$. Furthermore, a radial oscillation behavior is seen in the intensity profile across all LPF cases, as seen in the data from Fig.~\ref{fig:processed_data} and the model in Fig.~\ref{fig:Maximum Ring Radii Ranges}. The radial oscillation is a result of the diffraction of the high frequency information in the mode, and can be suppressed by applying a LPF. Applying a LPF generates a radial intensity gradient at $d_1$, which increases with a smaller filter radii due to the increase in $r_0$. Yet, a radial intensity gradient is the primary reason against using conventional optical vortices in cases where a uniform amplitude profile with an orbital phase is preferred. Therefore, practical use of a GV mode requires careful engineering of the optical system to ensure an appropriate mode is achieved, with sufficiently small radial oscillations, large $z_\mathrm{V}$, and small $r_0$.

\section*{Conclusion}
The limited aperture of physical optical circuits influences the vortex core size of a Gaussian vortex (GV) mode that is generated by imprinting a Gaussian beam with a helical phase. This paper explored the fundamental properties of GV beams in the low spatial frequency regime to characterise the effect of limited aperture in a typical optical setup. A simple relationship between the vortex core radius at the point of phase imprinting, the orbital quantum number of the optical vortex $\ell$, and the diameter of a low pass filter was established and found to be dependent on a calibration factor related to the specific experiment setup. Similar to the Rayleigh range for self-similar modes, the propagation distance over which vortex radius increased by $\sqrt{2}$ was measured, and was found to follow a power law with scaling factor that linearly relates to $\ell$.

\begin{appendices}

\section{\texorpdfstring{Modeling $\text{HyGG}^{\ell}_{p=-|\ell|}(r,\theta,z)$ from the Circular Beam Basis}{Modeling HyGG from the Circular Beam Basis}}\label{app:CiB_to_HyGG}

The Circular Beam (CiB) basis is the most general solution to the PWE for optical modes with intrinsic OAM, denoted in cylindrical coordinates $r$, $\theta$ and $z$, as detailed in \cite{circle_beams} and \cite{circle_beams2}. Omitting the Gouy phase shift, longitudinal phase accumulation, and wavefront curvature terms inherent to PWE solutions, CiBs are described by the expression
\begin{equation}\label{eq:2}
\begin{split}
&\text{CiB}^l_\gamma(r;q,\tilde{q})=\left(\frac{\Tilde{q}/\Tilde{q}_0}{q/q_0}\right)^{\frac{i\gamma}{2}}\left(\frac{ir^2}{2\chi^2}\right)^{-\frac{1}{2}}M_{i\gamma/2,l/2}\left(\frac{ir^2}{2\chi^2}\right)\left[\mathrm{G}(r,q)\mathrm{G}(r,\Tilde{q})\right]^{\frac{1}{2}}e^{i\ell\theta}
\end{split}
\end{equation}

\noindent where $q(z) = z + q_0$ and $\Tilde{q}(z) = z + \Tilde{q}_0$ are Seigman's complex beam parameter \cite{seigman}, and an integration constant governing the hypergeometric structure of the electric field, respectively \cite{circle_beams2}. Note that these are different than the two complex beam parameter formalism corresponding to the tangential and sagittal planes of a single mode \cite{independent_qs}. Eq. \ref{eq:2} also uses the fundamental Gaussian mode denoted by
\begin{equation}
\text{G}(r,q)=\left(1+\frac{z}{q_0}\right)^{-1}\exp\left(-i\frac{kr^2}{2q}\right)
\end{equation}

\noindent alongside the constant $\chi^2=\left(k\left(1/\Tilde{q} - 1/q\right)\right)^{-1}$ for wavenumber $k$, OAM quantum number $\ell$, and the solution to Whittaker differential equation, $M_{\delta,\mu}(\rho)$, which is a Whittaker function of argument $\rho$ and parameters $\delta$ and $\mu$ \cite{handbook_of_maths}. The choice of parameters $q_0$, $\Tilde{q}_0$, and $\gamma$ defines the particular class of CiB represented by Eq. \ref{eq:2}, including the Laguerre-Gauss modes in standard (sLG) \cite{seigman}, elegant \cite{generalised_GB_solutions}, and generalized \cite{cartesian_beams} form, alongside the Bessel-Gauss (BG) beams \cite{BG}, and the family of Hypergeometric (HyG) beams \cite{HyGG_fam}. To derive the Type I HyGG limit from Eq. \ref{eqn:GV}, we start with Eq. \ref{eq:2} and note that $p=-|\ell|\Rightarrow\gamma=-i$ then
\begin{equation}\label{enq:4}
\begin{split}
\text{HyGG}^{\ell}_{-\ell}(r,\theta,z) &= \text{CiB}^{\ell}_{-i}(r;q,\tilde{q}) \\
&= \left(\frac{\Tilde{q}/\Tilde{q}_0}{q/q_0}\right)^{\frac{1}{2}}\left(\frac{ir^2}{2\chi^2}\right)^{-\frac{1}{2}} M_{1/2,|l|/2}\left(\frac{ir^2}{2\chi^2}\right)\left[\mathrm{G}(r,q)\mathrm{G}(r,\Tilde{q})\right]^{\frac{1}{2}}e^{i\ell\theta}.\\
\end{split}
\end{equation}

\noindent Whittaker's function is related to a confluent hypergeometric function $_1F_1(a,b;z)$ for arbitrary a, b, and z \cite{handbook_of_maths} to a confluent hypergeometric function ${}_1F_1(a,b;z)$ for arbitrary $a$, $b$, and $z$
\begin{equation}
\begin{split}
M_{\delta,\mu}(z) = e^{-\frac{z}{2}}z^{\frac{1}{2}+\mu}{}_1F_1\left(\frac{1}{2}+\mu-\delta,1+2\mu;z\right)
\end{split}
\end{equation}

\noindent where $\delta = \frac{1}{2}$ and $\mu = \frac{|\ell|}{2}$. Substituting these into Eq. \ref{enq:4}
\begin{equation}
\begin{split}
\text{HyGG}^{\ell}_{-\ell}(r,\theta,z) &= \left(\frac{\Tilde{q}/\Tilde{q}_0}{q/q_0}\right)^{\frac{1}{2}}\exp\left(-\frac{ir^2}{4\chi^2}\right)\left(\frac{ir^2}{2\chi^2}\right)^{\frac{|\ell|}{2}} {}_1F_1\left(\frac{|\ell|}{2},1+|\ell|;\frac{ir^2}{2\chi^2}\right)\left[\mathrm{G}(r,q)\mathrm{G}(r,\Tilde{q})\right]^{\frac{1}{2}}e^{i\ell\theta}.\\
\end{split}
\end{equation}

\noindent Which can be simplified further by
\begin{equation}
\begin{split}
\exp\left(-\frac{ir^2}{4\chi^2}\right)\left[\mathrm{G}(r,q)\mathrm{G}(r,\Tilde{q})\right]^{\frac{1}{2}} &= \left(\left(1+\frac{z}{q_0}\right)\left(1+\frac{z}{\tilde q_{0}}\right)\right)^{-\frac{1}{2}}\exp\left(-\frac{ir^2}{4\chi^2} - \frac{ikr^2}{4q} - \frac{ikr^2}{4\tilde q}\right) \\
&= \left(\frac{q_0\tilde q_{0}}{q\tilde q}\right)^{\frac{1}{2}}\exp\left(-\frac{ir^2}{4\chi^2} - \frac{ikr^2}{4}\left(\frac{1}{q} + \frac{1}{\tilde q}\right)\right) \\
\end{split}
\end{equation}

\noindent such that
\begin{equation}\label{enq:5}
\begin{split}
\text{HyGG}^{\ell}_{-\ell}(r,\theta,z) &= \left(\frac{ir^2}{2\chi^2}\right)^{\frac{|\ell|}{2}} {}_1F_1\left(\frac{|\ell|}{2},1+|\ell|;\frac{ir^2}{2\chi^2}\right)\left(\frac{q_0}{q}\right)\exp\left(-\frac{ir^2}{4\chi^2} - \frac{ikr^2}{4}\left(\frac{1}{q} + \frac{1}{\tilde q}\right)\right)e^{i\ell\theta}.\\
\end{split}
\end{equation}

\noindent The simplest way to expand ${}_1F_1\left(\frac{|\ell|}{2},1+|\ell|;\frac{ir^2}{2\chi^2}\right)$ is via the power series expansion
\begin{equation}
\begin{split}
{}_1F_1\left(\frac{|\ell|}{2},1+|\ell|;\frac{ir^2}{2\chi^2}\right)
=\sum_{n=0}^\infty
\frac{\Gamma\left(\frac{|\ell|}{2}+n\right)\,\Gamma(1+|\ell|)}
     {\Gamma\left(\frac{|\ell|}{2}\right)\,\Gamma(1+|\ell|+n)}
\;\left(\frac{ir^2}{2\chi^2}\right)^n\frac{1}{n!}
\end{split}
\end{equation}

\noindent which takes the form of a Kummer function with Gamma function $\Gamma(x_1)$ for $x_1 \in \mathbb{C}\setminus\{0,-1,-2,\dots\}$. This can be written in closed‐form in terms of the modified Bessel function of the first kind $I_\nu(x)$ from Eq. 13.6.9 of \cite{NIST_maths}
\begin{equation}
\begin{split}
{}_1F_1\!\Bigl(\tfrac{|\ell|}{2},\,1+|\ell|;\,\frac{ir^2}{2\chi^2}\Bigr)
&=\frac{\Gamma(1+|\ell|)}{\Gamma\!\left(\tfrac{1+|\ell|}{2}\right)}
\exp\left(\frac{ir^2}{4\chi^2}\right)\left(\frac{ir^2}{4\chi^2}\right)^{\!\tfrac{1-|\ell|}{2}}I_{\tfrac{|\ell|-1}{2}}\!\left(\frac{ir^2}{4\chi^2}\right) \\
\end{split}
\end{equation}

\noindent for $x,\nu \in \mathbb{C}$. This is often much more compact (and numerically robust) than the infinite series. Substituting this into Eq. \ref{enq:5}
\begin{equation}\label{enq:7}
\begin{split}
\text{HyGG}^{\ell}_{-\ell}(r,\theta,z) &= \left(\frac{ir^2}{4\chi^2}\right)^{\!\tfrac{1}{2}}I_{\tfrac{|\ell|-1}{2}}\!\left(\frac{ir^2}{4\chi^2}\right)\left(\frac{q_0}{q}\right)\exp\left(- \frac{ikr^2}{4}\left(\frac{1}{q} + \frac{1}{\tilde q}\right)\right)e^{i\ell\theta}\\
\end{split}
\end{equation}

\noindent which is the simplified analytical expression for a HyGG mode with $p=-|\ell|$ in the CiB's notation from \cite{circle_beams}, where the constants have been incorporated into $C$.

\section{Deriving the Gaussian Vortex Expression}\label{app:GV_derivation}

By considering the HyGG expression from Eq. \ref{enq:7} in the limit of $z\rightarrow d_1$ then $\chi^2 \rightarrow 0$. Therefore, the argument of the Bessel function becomes large, and Eq. 10.40.1 from \cite{NIST_maths} tells us 
\begin{equation}
\begin{split}
I_{\nu}(x)\sim
\frac{e^{x}}{\sqrt{2\pi\,x}}
\sum_{k=0}^{\infty}(-1)^{k}\,\frac{a_{k}(\nu)}{x^{k}}
\quad
\bigl(x\to\infty,\;|\arg x|<\tfrac{\pi}{2}\bigr)
\end{split}
\end{equation}
\noindent for Pochhammer symbol $a_k(\nu)$. So a first order approximation is
\begin{equation}
\begin{split}
I_{\frac{|l|-1}{2}}\left(\frac{ir^2}{4\chi^2}\right)&\sim\left(\frac{2\chi^2}{i\pi r^2}\right)^{\frac{1}{2}}\exp\left(\frac{ir^2}{4\chi^2}\right) \\
\end{split}
\end{equation}

\noindent which we can substitute back into Eq. \ref{enq:7} such that,
\begin{equation}
\begin{split}
\lim_{z \to d_1}\text{HyGG}^{\ell}_{p=-|\ell|}(r,\theta,z) &\sim \left(\frac{ir^2}{4\chi^2}\right)^{\!\tfrac{1}{2}}\left(\frac{2\chi^2}{i\pi r^2}\right)^{\frac{1}{2}}\exp\left(\frac{ir^2}{4\chi^2}\right)\left(\frac{q_0}{q}\right)\exp\left(- \frac{ikr^2}{4}\left(\frac{1}{q} + \frac{1}{\tilde q}\right)\right)e^{i\ell\theta}\\
&= \left(\frac{1}{2\pi}\right)^{\frac{1}{2}}\left(\frac{q_0}{q}\right)\exp\left(\frac{ikr^2}{2q}\right)e^{i\ell\theta}\\
&\propto\text{G}(r,d_1-d_0+q_0)e^{i\ell\theta}\\
\end{split}
\end{equation}
which is a standard Gaussian mode, $\text{G}(r,q)$, with an orbital phase factor.

\section{Applying a Low Pass Filter to a Gaussian Vortex Mode}\label{app:low pass filter derivation}

Applying a low pass filter to the GV from Eq. \ref{eqn:GV}, we can rewrite the GV mode as
\begin{equation}
\begin{split}
\lim_{z \to d_1}\text{HyGG}^{\ell}_{p=-|\ell|}(r,\theta,z) &= \text{GV}^{\ell}(r,\theta,d_1)\\
&= C\exp\left(-\frac{w_0^2}{r^2} + i\frac{2d_1}{kr^2} + i\ell\theta\right) \\
\end{split}
\end{equation}
for constants $C$. We then perform a Fourier transform on $\text{GV}^l(r,\theta,d_1)$ into polar frequency coordinates $(k_{r},\phi)$ where
\begin{equation}
    \widetilde F(k_{r},\phi)
=\mathcal{F}\{\text{GV}^{\ell}(r,\theta,d_1))\}(k_r,\phi)
=\int_{0}^{\infty}\!\int_{0}^{2\pi}
Ce^{-w_0^2/r^2 + i2d_1/kr^2 + i\ell\theta}e^{-i\,k_{r}r\cos(\theta-\phi)}r\mathrm{d}\theta\mathrm{d}r.
\end{equation}
Using Eq.s 9.1.41 and 9.1.42 from \cite{handbook_of_maths}, we get the identity
\begin{equation}\label{eqn:the_identity}
\begin{split}
\int_{0}^{2\pi}e^{i\ell\theta}\,e^{-i\,k_{r}r\cos(\theta-\phi)}\mathrm{d}\theta
=2\pi\,i^{\,\ell}\,e^{i\ell\phi}\,J_{\ell}(k_{r}r),
\end{split}
\end{equation}
where $J_{\ell}$ is the Bessel function of the first kind of order $\ell$. We are left with the Hankel transform of order \(\ell\) \cite{Hankel_transform}:
\begin{equation}
\begin{split}
\widetilde F(k_{r},\phi)
=2\pi\,C\,i^{\,\ell}\,e^{\,i\ell\phi}
\int_{0}^{\infty}
\exp\!\Bigl(-\frac{w_{0}^{2}}{r^{2}}+i\frac{2\,d_{1}}{k\,r^{2}}\Bigr)
\,J_{\ell}(k_{r}r)r\mathrm{d}r.
\end{split}
\end{equation}
Defining $\beta = w_{0}^{2} \;-\; i\,\frac{2\,d_{1}}{k}$  one finds
\begin{equation}\label{eqn:Fourier_transformed}
\begin{split}
\int_{0}^{\infty}r\,e^{-\beta/r^{2}}\,J_{\ell}(k_{r}r)\,dr
&=\frac{\beta^{\tfrac{\ell+2}{2}}}{2}\;k_{r}^{\ell}\;
U\!\Bigl(\tfrac{\ell+2}{2},\,\ell+1,\;\tfrac{\beta\,k_{r}^{2}}{4}\Bigr) \\
\Rightarrow F(k_{r},\phi)
&=\;\pi\,C\;i^{\,\ell}\;e^{\,i\ell\phi}\;\,
\beta^{\tfrac{\ell+2}{2}}\;k_{r}^{\ell}\;
U\!\Bigl(\tfrac{\ell+2}{2},\,\ell+1,\;\tfrac{\beta\,k_{r}^{2}}{4}\Bigr),\\
\end{split}
\end{equation}
\noindent where $U$ is Tricomi’s confluent hypergeometric function \cite{Tricomi1955}. To apply a low pass filter, we define the transfer function
\begin{equation}
H(k_r)=
\begin{cases}
1, & 0 \le k_r \le k_c,\\
0, & k_r > k_c
\end{cases}
\end{equation}
\noindent where $k_c$ is the cutoff frequency. Instead of picturing this as a GV mode propagating into the near field where it is Fourier filtered by optics of limited aperture, we use a simpler approach by making Fourier space located on the focal plane between the two lenses of a 4f imaging system, as we do experimentally in Fig. \ref{fig:setup}. In this case, $k_c$ is related to the diameter $D$ of an aperture in real space by $k_c = kD/2f$ for wavenumber $k$ and objective lens focal length $f$ \cite{fourier_optics}. Applying the low pass filter
\begin{equation}
\widetilde F_{\rm filt}(k_r,\phi)
= \widetilde F(k_r,\phi)\,H(k_r)
= \pi\,C\,i^{\ell}\,e^{\,i\ell\phi}\,
\beta^{\tfrac{\ell+2}{2}}\,k_{r}^{\ell}\,
U\!\Bigl(\tfrac{\ell+2}{2},\,\ell+1,\;\tfrac{\beta\,k_{r}^{2}}{4}\Bigr)
\;H(k_r).
\end{equation}
\noindent To see the effect in real space, we then perform the inverse Fourier transform
\begin{equation}
F_{\rm filt}(r,\theta)
=\int_{0}^{2\pi}\!\int_{0}^{\infty}
\widetilde F_{\rm filt}(k_r,\phi)\,
e^{\,i\,k_r r\cos(\theta-\phi)}k_r\mathrm{d}k_r\mathrm{d}\phi
\end{equation}
which reduces to
\begin{equation}
\begin{split}
F_{\rm filt}(r,\theta)
&=2\pi\,C\,i^{\ell}e^{\,i\ell\theta}\beta^{\tfrac{\ell+2}{2}}
\int_{0}^{k_c}k_{r}^{\ell+1}U\!\Bigl(\tfrac{\ell+2}{2},\,\ell+1,\;\tfrac{\beta\,k_{r}^{2}}{4}\Bigr)
\,J_{\ell}(k_r r)\mathrm{d}k_r \\
&=C\int_{0}^{k_c}k_{r}^{\ell+1}U\!\Bigl(\tfrac{\ell+2}{2},\,\ell+1,\;\tfrac{\beta\,k_{r}^{2}}{4}\Bigr)
\,J_{\ell}(k_r r)\mathrm{d}k_r \\
\end{split}
\end{equation}
when the identity from Eq. \ref{eqn:the_identity} is used, and $C$ contains all constants.

\section{Analytically Relating $r_0$ to $\ell$ at $d_1$}\label{app: solving A_l}
Rewriting Eq. \ref{eqn:derivative} in our dimensionless coordinates, we define the parameter $\alpha \in \mathbb{R}^+$ such that when $r = r_0 \Rightarrow u = \alpha = k_cr$, meaning
\begin{equation}\label{eqn:derivation_final}
\begin{split}
\left.\frac{\partial}{\partial r}\big|\text{GV}_{\rm filt}(r,\theta)\big|\right|_{r=r_0}=0 &\Rightarrow \frac{\mathrm d}{\mathrm d u}\left[u^{-(\ell+2)}A(u)\right]_{u=\alpha}=0. \\
\end{split}
\end{equation}
for $A(u)$ from Eq. \ref{eqn:A_l}. For most optical circuits, $|\beta|^2k_c^2 \gg 1$, meaning we can define $\eta = (\ell + 2)/2$, $\zeta = \ell + 1$, and $\delta = \beta k_c^2 s^2/u^2$ and rewrite $U$ under the asymptotic limit \cite{Tricomi1955}
\begin{equation}
\begin{split}
    U(\eta,\zeta,\delta) &= \delta^{-\eta}\left(1 - \frac{\eta(\eta-\zeta+1)}{\delta} + \mathcal O\left(\delta^{-2}\right)\right) \\
    &= \delta^{-\eta}\left(1 - \frac{1}{\delta}\left(1 - \frac{\ell^2}{4}\right) + \mathcal O\left(\delta^{-2}\right)\right). \\
\end{split}
\end{equation}
\noindent Therefore, to leading order
\begin{equation}
\begin{split}
A(u) &\approx \left(\frac{1}{\beta k_c^2}\right)^{\frac{\ell + 2}{2}} u^{\ell + 2}\int^u_0 \frac{J_\ell(s)}{s}\mathrm{d}s \\
\end{split}
\end{equation}
meaning Eq. \ref{eqn:derivation_final} becomes
\begin{equation}
\begin{split}
\frac{\mathrm d}{\mathrm d u}\left[\left(\beta k_c\right)^{\ell + 2}\int^u_0 \frac{J_\ell(s)}{s}\mathrm{d}s\right]_{u=\alpha} = \frac{J_\ell(\alpha)}{\alpha} \approx 0
\end{split}
\end{equation}
to leading order, which means at $r_0$, $\alpha \approx j_{\ell,1}$ where $j_{\ell,1}$ is the first positive root of the Bessel function, whose values are found in Tables 9.5-9.6 of \cite{handbook_of_maths}. Therefore, we can approximate
\begin{equation}
    r_0\approx\frac{j_{\ell,1}}{k_c}.
\end{equation} 

\end{appendices}

\section*{Funding} 
Australian Research Council Grants No. FT210100809 and No. LP190100621.

\section*{Acknowledgments}
The authors would like to thank Assoc. Prof. John Debs, Prof. Cass Sackett, Dr. Michael Larson, and Dr. Eric Imhof for their insightful discussions contributing to this work. Additionally, the authors were supported through Australian Research Council Grants No. FT210100809 (S.A.H.) and No. LP190100621 (J.E., R.J.T., and S.L.).

\section*{Disclosures} The authors declare no conflicts of interest.

\section*{Data Availability} Data underlying the results presented in this paper are not publicly available at this time but may be obtained from the authors upon reasonable request.

\bibliography{biblio}

\end{document}